\documentclass[sigconf]{acmart}

\usepackage{amsmath}
\usepackage{array}
\usepackage{booktabs}
\usepackage{graphicx}
\usepackage{makecell}
\usepackage{multirow}
\usepackage{needspace}
\usepackage{subcaption}
\usepackage{tabularx}
\usepackage{latexml}
\usepackage{tikz}
\usepackage{pgfplots}
\usepackage{xcolor}
\newcommand{\methodname}{DOiT}
\pgfplotsset{compat=1.18}
\usepgfplotslibrary{groupplots}
\usetikzlibrary{arrows.meta,calc,positioning}

\copyrightyear{2026}
\acmYear{2026}
\setcopyright{cc}
\setcctype{by}
\acmConference[ICAIF '26]{7th ACM International Conference on AI in Finance}{November 14--17, 2026}{Milan, AA, Italy}
\acmBooktitle{7th ACM International Conference on AI in Finance (ICAIF '26), November 14--17, 2026, Milan, AA, Italy}
\acmDOI{10.1145/3858063.3858842}
\acmISBN{979-8-4007-3087-0/2026/11}
\begin{document}

\title{\methodname: Better Money-Flow Ranking Can Coexist with Worse Account Selection in Anti-Money Laundering Review}
\thanks{Author-prepared version of the ICAIF '26 paper. The ACM version of record is available at \url{https://doi.org/10.1145/3858063.3858842}.}

\author{Gwangui Hong}
\orcid{0009-0003-9160-6699}
\affiliation{\institution{KAIST}\city{Daejeon}\country{South Korea}}
\affiliation{\institution{KFTC}\city{Seoul}\country{South Korea}}
\email{recordame@kaist.ac.kr}
\email{recordame@kftc.or.kr}

\author{Seonkyu Lim}
\orcid{0000-0003-4904-7144}
\iflatexml
  \authornote{Corresponding author.}
\else
  \authornote{Corresponding authors.}
\fi
\affiliation{\institution{KFTC}\city{Seoul}\country{South Korea}}
\email{sklim@kftc.or.kr}

\author{Min-Soo Kim}
\orcid{0000-0002-5065-0226}
\iflatexml
  \authornote{Corresponding author.}
\else
  \authornotemark[1]
\fi
\affiliation{\institution{KAIST}\city{Daejeon}\country{South Korea}}
\email{minsoo.k@kaist.ac.kr}

\begin{abstract}
Anti-money laundering (AML) review selects suspicious accounts and reconstructs transfers. Downstream graph neural networks (GNNs) rank money-flow evidence but may also reorder the account queue. \methodname{} separates these decisions at equal review capacity: account selection determines where graph investigation starts. Across AMLWorld and AMLNet, GNN reordering harms higher-yield upstream orders but helps lower-yield ones. On AMLWorld, this harm coexists with higher transaction average precision (AP) in dimension-matched, per-run decoders, demonstrating separable ranking abilities. At small capacities, positive neighbors can displace a denser upstream prefix and lower yield; replacement costs fall as that prefix is exhausted, allowing the preferred order to reverse. A restricted-access synthetic benchmark informed by real transaction patterns reproduces this separation and validation-based upstream preservation. Capacity-matched combination protects upstream ordering while retaining GNN money-flow analysis, admitting GNN queue influence only where held-out evidence supports it.
\par\noindent Source code: \url{https://github.com/recordame/DOIT}.
\end{abstract}



\iflatexml
  \maketitle
\else
  \makeatletter
  \let\doit@fnsymbol\@fnsymbol
  \def\@fnsymbol#1{\ifnum#1=1\relax\doit@fnsymbol{2}\else\doit@fnsymbol{#1}\fi}
  \maketitle
  \let\@fnsymbol\doit@fnsymbol
  \makeatother
\fi

\section{Introduction}
\label{sec:introduction}

Money laundering conceals the origin of criminal proceeds. Anti-money laundering (AML) review identifies suspicious activity~\cite{basel2020soundaml,oztas2024transactionmonitoring} and assembles fund-flow evidence~\cite{ffiec_sar_quality}. Under limited resources, two complementary decisions follow: account selection determines where investigation starts; money-flow analysis identifies supporting transfers, connected accounts, and paths.

Models support both decisions. A selector orders candidates from monitoring alerts, account features, and transaction histories; graph-based AML methods analyze accounts, transactions, or suspicious subgraphs, depending on the task~\cite{cardoso2022laundrograph,li2020flowscope,cheng2023groupaml}. In the workflow studied here, a downstream GNN supplies money-flow evidence together with account scores. An AML pilot similarly connects monitoring-based risk assessment to fund-flow investigation before a final decision~\cite{jfsa2026fintechhub}. Because review begins from a finite queue, GNN account scores may also reorder the queue analysts actually work. We use $K$ for the number of accounts that can be reviewed and $\mathrm{Pr}@K$ for the positive rate among the first $K$ accounts. Figure~\ref{fig:workflow} compares the two orders at the same $K$. This handoff is convenient, but the scores answer different questions.

The selector chooses accounts for review; the GNN identifies suspicious account context, transfers, and paths. If its scores reorder the final queue, accounts raised by relational context displace upstream choices. This is score-based reranking; no review slots are reserved for neighbors. Studies of graph representation, suspicious flows, and laundering groups evaluate performance within their respective tasks~\cite{cardoso2022laundrograph,li2020flowscope,cheng2023groupaml}. The handoff considered here additionally asks whether GNN reranking preserves an independent upstream account list at the capacity where it is worked. \methodname{} (\emph{Downstream Ordering isn't Transaction Ranking}) asks when GNN account scores should change an existing review queue, not whether the GNN learns useful graph structure.

Answering this requires three controls: measure transaction evidence and account yield separately because they rank different objects; hold the graph builder, trained GNN, and frozen outputs fixed because the selector changes the opened surface; and compare the two orders across planned review capacities. Staffing, not the model, sets review capacity, so the answer is not a single number.

Figure~\ref{fig:gnn-paradox} previews the central failure mode: a GNN can
improve ranking from an independently trained transaction (Tx) decoder while its account scores
replace positives in the upstream queue.

\ifdefined\KoreanPaper
  \def\ParStageOne{1. 선별}
  \def\ParStageTwo{2. 다운스트림}
  \def\ParStageThree{3. 제한 검토}
  \def\ParDenseGraph{선별 목록이 연 계좌 표면}
  \def\ParGNNRep{고정\\GNN}
  \def\ParRestrict{제한}
  \def\ParTxDecoder{거래\\디코더}
  \def\ParAccountScore{계좌\\점수}
  \def\ParTxUp{거래 AP $\uparrow$}
  \def\ParRawAccount{GNN 계좌 원순위}
  \def\ParDisplaced{선별 단계 양성 계좌가 밀려남}
  \def\ParCaption{AMLWorld의 CLeAR--GraphSAGE 사례($K=100$, 10회 평균). 거래 AP는 시드별 별도 디코더, 계좌 순위는 고정된 GNN 계좌 점수로 계산한다. AP는 평균정밀도다.}
  \def\ParDescription{고수율 CLeAR 선별 순위가 양성 밀도가 높은 그래프와 상위 100개 양성률 0.89를 만든다. GraphSAGE는 거래 순위를 개선하지만 선별 계좌와 연결 계좌의 순서도 바꾼다. 그 결과 선별 단계 양성 계좌가 밀려나 최종 상위 100개 계좌의 양성률은 0.82가 된다.}
\else
  \def\ParStageOne{1. UPSTREAM}
  \def\ParStageTwo{2. DOWNSTREAM}
  \def\ParStageThree{3. LIMITED REVIEW}
  \def\ParDenseGraph{surface opened by upstream queue}
  \def\ParGNNRep{frozen GNN}
  \def\ParRestrict{restrict}
  \def\ParTxDecoder{Tx decoder}
  \def\ParAccountScore{account score}
  \def\ParTxUp{transaction AP $\uparrow$}
  \def\ParRawAccount{raw GNN account order}
  \def\ParDisplaced{upstream positives displaced}
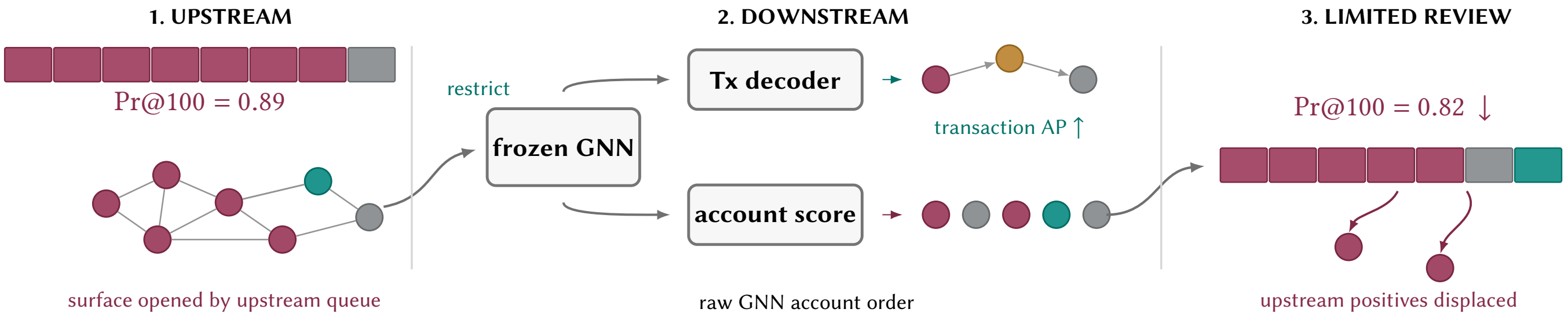
  \def\ParCaption{AMLWorld CLeAR--GraphSAGE example at $K=100$ (ten-run means). Transaction average precision (AP) uses independently fitted per-seed decoders; frozen GNN scores order accounts.}
  \def\ParDescription{Three-stage diagram. A higher-yield CLeAR upstream order supplies a positive-dense graph and a top-100 positive rate of 0.89. GraphSAGE improves transaction ranking but also reorders upstream and connected accounts. The final top-100 account list has positive rate 0.82 because positive upstream accounts are displaced.}
\fi
\ifdefined\KoreanPaper
  \begin{figure}[!t]
  \def\ParFigureWidth{\columnwidth}
  \def\ParHorizontalScale{1}
  \def\ParSlotWidth{2.10mm}
  \def\ParScoreWidth{.60cm}
  \def\ParHeadWidth{.66cm}
\else
  \begin{figure*}[!t]
  \def\ParFigureWidth{\textwidth}
  \def\ParHorizontalScale{1.8}
  \def\ParSlotWidth{3.78mm}
  \def\ParScoreWidth{1.20cm}
  \def\ParHeadWidth{1.40cm}
\fi
  \centering
  \resizebox{\ParFigureWidth}{!}{%
  \begin{tikzpicture}[
      xscale=\ParHorizontalScale,
      font=\sffamily\scriptsize,
      >=Latex,
      score/.style={rounded corners=2pt, draw=black!54, fill=black!3,
        line width=.65pt, minimum width=\ParScoreWidth, minimum height=.62cm,
        inner xsep=1.2pt, font=\sffamily\bfseries\scriptsize, align=center},
      dot/.style={circle, draw=#1!80!black, fill=#1!88, line width=.38pt,
        minimum size=2.15mm, inner sep=0pt},
      slot/.style={rounded corners=.42pt, draw=#1!82!black, fill=#1!86,
        line width=.32pt, minimum width=\ParSlotWidth, minimum height=2.75mm,
        inner sep=0pt},
      flow/.style={-{Latex[length=1.15mm,width=.88mm]}, draw=black!56,
        line width=.56pt, shorten <=.8mm, shorten >=.8mm},
      branchflow/.style={-{Latex[length=1.08mm,width=.82mm]}, draw=black!56,
        line width=.54pt, shorten <=1.25mm, shorten >=1.35mm},
      outputflow/.style={-{Latex[length=1.00mm,width=.76mm]}, draw=black!56,
        line width=.52pt, shorten <=1.45mm, shorten >=1.45mm},
      midarrow/.style={-{Latex[length=.95mm,width=.72mm]}, draw=black!56,
        line width=.26pt},
      relation/.style={draw=black!42, line width=.34pt},
      head/.style={font=\sffamily\bfseries\fontsize{5.2}{5.7}\selectfont,
        align=center},
      note/.style={font=\sffamily\tiny, text=black, align=center},
      metric/.style={font=\sffamily\bfseries\scriptsize, align=center}
  ]
  \definecolor{parawine}{RGB}{150,48,83}
  \definecolor{parateal}{RGB}{0,135,125}
  \definecolor{paraorange}{RGB}{184,126,38}
  \definecolor{paragray}{RGB}{128,132,134}

  \node[head] at (.96,2.42) {\ParStageOne};
  \foreach \x/\col in {
      .10/parawine,.32/parawine,.54/parawine,.76/parawine,
      .98/parawine,1.20/parawine,1.42/parawine,1.64/paragray}
    {\node[slot=\col] at (\x,2.03) {};}
  \node[metric, text=parawine!92!black] at (.87,1.72)
    {$\mathrm{Pr}@100=0.89$};

  \node[dot=parawine] (u1) at (.45,.91) {};
  \node[dot=parawine] (u2) at (.72,1.14) {};
  \node[dot=parawine] (u3) at (1.00,.92) {};
  \node[dot=parawine] (u4) at (.68,.62) {};
  \node[dot=parawine] (u5) at (1.24,.62) {};
  \node[dot=parateal] (u6) at (1.40,1.09) {};
  \node[dot=paragray] (u7) at (1.63,.80) {};
  \draw[relation] (u1)--(u2)--(u3)--(u6)--(u7);
  \draw[relation] (u1)--(u4)--(u3)--(u5)--(u7);
  \draw[relation] (u2)--(u4) (u3)--(u6) (u4)--(u5);
  \node[note, text=parawine!88!black] at (.98,.12)
    {\ParDenseGraph};

  \node[head] at (3.62,2.42) {\ParStageTwo};
  \node[score] (gnn) at (2.50,1.365) {\ParGNNRep};
  \draw[flow] (1.65,.87) to[out=18,in=205] (gnn.west);
  \node[note,text=parateal!85!black] at (2.12,1.84) {\ParRestrict};

  \node[score,minimum width=\ParHeadWidth,minimum height=.48cm] (txdec) at (3.45,1.91) {\ParTxDecoder};
  \node[score,minimum width=\ParHeadWidth,minimum height=.48cm] (acct) at (3.45,.82) {\ParAccountScore};
  \draw[branchflow] (gnn.north) to[out=90,in=180] (txdec.west);
  \draw[branchflow] (gnn.south) to[out=-90,in=180] (acct.west);

  \node[dot=parawine] (p1) at (4.17,1.91) {};
  \node[dot=paraorange] (p2) at (4.50,2.08) {};
  \node[dot=paragray] (p3) at (4.83,1.91) {};
  \draw[midarrow,draw=parateal!84!black]
    (3.93,1.91) -- (4.02,1.91);
  \draw[relation,-{Latex[length=.9mm,width=.68mm]}] (p1)--(p2);
  \draw[relation,-{Latex[length=.9mm,width=.68mm]}] (p2)--(p3);
  \node[note, text=parateal!85!black] at (4.50,1.52)
    {\ParTxUp};

  \node[dot=parawine] (a1) at (4.17,.82) {};
  \foreach \x/\y/\col in {
      4.35/.82/paragray,4.53/.82/parawine,
      4.71/.82/parateal,4.89/.82/paragray}
    {\node[dot=\col] at (\x,\y) {};}
  \draw[midarrow,draw=parawine!82!black]
    (3.93,.82) -- (4.02,.82);
  \node[note] at (3.59,.12) {\ParRawAccount};

  \node[head] at (6.28,2.42) {\ParStageThree};
  \draw[flow] (4.89,.82) to[out=0,in=190] (5.42,1.22);
  \foreach \x/\col in {
      5.55/parawine,5.77/parawine,5.99/parawine,
      6.21/parawine,6.43/parawine,6.65/paragray,6.87/parateal}
    {\node[slot=\col] at (\x,1.22) {};}
  \node[metric, text=parawine!92!black] at (6.22,1.68)
    {$\mathrm{Pr}@100=0.82\,\downarrow$};

  \node[dot=parawine] (lost1) at (6.02,.56) {};
  \node[dot=parawine] (lost2) at (6.43,.39) {};
  \draw[-{Latex[length=1.00mm,width=.76mm]}, draw=parawine!86!black,
        line width=.50pt] (6.24,1.01) to[out=-105,in=75] (lost1.north);
  \draw[-{Latex[length=1.00mm,width=.76mm]}, draw=parawine!86!black,
        line width=.50pt] (6.55,1.01) to[out=-70,in=80] (lost2.north);
  \node[note, text=parawine!90!black] at (6.20,.12)
    {\ParDisplaced};

  \draw[black!16,line width=.45pt] (1.82,.38)--(1.82,2.18);
  \draw[black!16,line width=.45pt] (5.18,.38)--(5.18,2.18);
  \end{tikzpicture}%
  }
  \caption{\ParCaption}
  \Description{\ParDescription}
  \label{fig:gnn-paradox}
\ifdefined\KoreanPaper
  \end{figure}
\else
  \end{figure*}
\fi

We evaluate this question on AMLWorld and AMLNet by sweeping review capacity, comparing account orders, and decomposing added and removed accounts. GNN reranking harms a higher-yield order but helps a lower-yield one, and the better order reverses only after the higher-yield order has spent its positive-dense prefix. We call such a later reversal a \emph{crossover}. One replacement balance explains both outcomes; capacity-matched combination protects that prefix without discarding graph evidence.

\Needspace{5\baselineskip}
This paper makes three contributions.
\begin{itemize}
  \item We introduce \methodname, a capacity-matched methodology separating money-flow ranking from direct account selection.
  \item We show that tracing money and selecting accounts are separable abilities: the two results can move in opposite directions, and the preferred account order depends on review capacity rather than being a fixed property of a model.
  \item We explain gains and losses through the balance of added and displaced positives, control queue influence with capacity-matched validation, and test transfer on a restricted-access synthetic benchmark reflecting real transaction patterns.
\end{itemize}

\input{resources/fig_01_selection_flow_review_pipeline.tex}

\section{Related Work}
\label{sec:related-work}

\subsection{Account Selection and Graph Construction}
\label{sec:related-account-selection}

AML monitoring uses red flags and contextual indicators to identify transactions or accounts for further review~\cite{ffiec_redflags,fintrac_indicators2025}. Prior work qualifies AML alarms with deep learning~\cite{jensen2023alarms}, detects laundering agents from transaction streams~\cite{monlad2022}, or learns account representations from transaction sequences~\cite{huang2024clear}. These methods provide review scores or candidates at their respective units. Professional laundering can distribute roles across networks, so an ordered account list is a starting point rather than a complete case~\cite{fatf2018pml}. In our workflow, \emph{upstream account selection} orders accounts for review.

Operational AML work can then examine fund flows and related accounts before the institution makes its final decision~\cite{jfsa2026fintechhub}. We study how this deeper graph analysis changes the upstream order when it also supplies account scores.

Selected accounts can seed a transaction graph for deeper analysis. Clue-conditioned group discovery and transaction-seeded subnetwork discovery motivate local graph investigation~\cite{wang2026clue2group,chai2023amap}. Other systems identify suspicious flows or contextualize prioritized transactions as graph-based cases~\cite{li2020flowscope,silva2026graphprioritization}. These tasks differ from revising an existing account order with added counterparties.

\subsection{Graph Learning for Money-Flow Analysis}
\label{sec:related-graph-learning}

Financial transactions form graphs in which accounts are connected by transfers. The two architectures evaluated here cover neighborhood aggregation and directed multigraph message passing~\cite{hamilton2017sage,egressy2024directedmultigraph}. AML graph methods support customer and transaction representations~\cite{cardoso2022laundrograph}, laundering-group detection~\cite{cheng2023groupaml}, and suspicious flow or subgraph discovery~\cite{li2020flowscope,chai2023amap}. Multi-level anomaly models further connect node, edge, and graph decisions~\cite{lin2024unigad}, while supervised node-anomaly detector rankings depend on evaluation protocol and non-GNN controls~\cite{tang2023gadbench}. Account classification asks whether an account receives the correct label; flow discovery asks whether connected transactions or paths show how funds moved. Our evaluation separates transaction ranking from account selection: better transaction or path ranking need not improve account yield.

\subsection{Account Ordering Under Limited Capacity}
\label{sec:related-capacity}

When positive cases are rare and labels are costly, the first part of a ranked list matters more than performance averaged over the full population. Active search studies the related objective of finding many positives with limited inspections~\cite{garnett2012bayesian}. AML institutions must also examine supporting transactions and document the factual basis before making a suspicious-transaction decision~\cite{ffiec_sar_quality,fintrac_str_guidance}. Recent work further shows that transaction-level and actor-level AML queues need not agree~\cite{malik2026queuegranularity}.

Outside AML, calibration improves probabilistic interpretation, and prior-probability adjustment accounts for changing class priors~\cite{guo2017calibration,saerens2002prior}. Neither procedure directly optimizes positive-account yield at the intended review capacity.

The closest AML studies address adjacent but different handoffs. Malik~\cite{malik2026queuegranularity} compares transaction-projected and native actor queues under a common actor review unit and evaluates a validation-tuned hybrid. Silva and Lima de Castro~\cite{silva2026graphprioritization} form graph-based cases from prioritized transactions, while Clue2Group~\cite{wang2026clue2group} reconstructs groups from initial clues. \methodname{} studies a different handoff: an existing upstream account order opens a neighbor-expanded surface for frozen downstream GNN scores. It separates transaction ranking from account yield, compares account orders at the same $K$, attributes gains and losses to account replacement, and controls queue influence through capacity-matched validation.

\section{\methodname{} Methodology}
\label{sec:problem-setting}

\methodname{} follows the operational handoff in Figure~\ref{fig:workflow}, while separating model training from capacity-specific evaluation. An upstream selector orders candidate accounts. A GNN is trained once on the fixed account-disjoint transaction graph and produces frozen account scores and hidden representations. At each review capacity, the upstream prefix opens a local surface by adding stored one-hop neighbors, and \methodname{} restricts the frozen downstream account order to that surface. The transaction audit uses the corresponding surface at its audited capacity. We compare both orders at the same capacity, tracing accounts retained, removed, or added.

\subsection{Upstream Account Selection}
\label{sec:upstream-selection}

Let $\mathcal{C}$ be a fixed candidate pool produced from available monitoring signals. Each test account $v$ has a binary evaluation label $y(v)\in\{0,1\}$; label $1$ denotes a positive account. Labels are used only for evaluation. The candidate pool, trained models, and score vectors remain fixed while the number of reviewed accounts changes.

An upstream selector scores each account in $\mathcal{C}$ from alerts, account features, and transaction history. Sorting these scores gives $U=(u_1,\ldots,u_{|\mathcal{C}|})$, and $U_K=\{u_1,\ldots,u_K\}$ is the upstream list at review capacity $K$. This list answers which $K$ accounts would be reviewed if the upstream order were used alone. It also supplies the starting accounts for downstream graph construction.

\subsection{Graph Surface and Downstream Analysis}
\label{sec:graph-construction}

Let $G=(V,E)$ be the fixed account-transaction graph used to train the GNN, with $\mathcal{C}\subseteq V$. The prepared graph stores at most five one-hop neighbors per account under the same sampling rule used during GNN training. For each capacity $K$, the surface builder starts from $U_K$ and adds those stored counterparties. Writing this bounded neighborhood as $N_5(u)$, the resulting account set is
\[
  S_K=U_K\cup\left(\bigcup_{u\in U_K}N_5(u)\right).
\]
Thus $S_K$ preserves the $K$ selected accounts and adds their stored neighbors as reproducible local context. Both selectors use the same rule; only the accounts opening the surface differ.

The GNN's account score $s_D(v)$ and hidden representation are computed once on the fixed graph, then restricted to $S_K$. This fixed-model protocol isolates queue control from adaptation: changes across $K$ arise from the surface and order, not refitting. Surface-specific retraining or message-passing inference is a distinct protocol. Writing $\operatorname{TopK}_K$ for the $K$ highest-scoring accounts, $D_K=\operatorname{TopK}_K\{s_D(v):v\in S_K\}$. A new neighbor enters only if it ranks ahead of an upstream top-$K$ account. Expansion creates candidates; ordering allocates the fixed review capacity.

We measure money-flow ranking with two audits. The primary control fits one logistic transaction decoder per selector--GNN pair and evaluates its test surface once. Both sides begin with one scalar account score and form the same four endpoint features for transaction $(i,j)$: $[s(i),s(j),|s(i)-s(j)|,s(i)s(j)]$. Decoder structure, standardization, feature dimension, transactions, and labels are identical within each pair; only the score source differs. Transactions enter only when both endpoints lie in $S_K$. Directed two-transaction paths are too sparse to supply both validation classes in every run, so they are excluded from the independent-decoder claim. A secondary representation audit retains hidden-state and score-induced transaction and path orders as diagnostics. Both audits rank relations rather than accounts in a capacity-limited review queue.

\subsection{Capacity-Matched Account Comparison}
\label{sec:review-capacity}

The subscripts $U$ and $D$ denote the upstream and downstream positive rates, respectively:
\[
\begin{aligned}
  \mathrm{Pr}_{U}@K&=\frac{1}{K}\sum_{v\in U_K}y(v),
  &\mathrm{Pr}_{D}@K&=\frac{1}{K}\sum_{v\in D_K}y(v).
\end{aligned}
\]
At every evaluated $K$, investigators can review exactly $K$ accounts from either order. This shared denominator makes the comparison operationally direct: a change in positive rate can only arise because the downstream order replaces accounts. Better transaction or path ranking need not imply better account selection, because the two measurements rank different objects.

At each $K$, let $O_K$, $R_K$, and $A_K$ denote the common, upstream-only, and downstream-only accounts, respectively:
\[
\begin{aligned}
  O_K &= U_K \cap D_K && \text{(common accounts)},\\
  R_K &= U_K \setminus D_K && \text{(upstream-only accounts)},\\
  A_K &= D_K \setminus U_K && \text{(downstream-only accounts)}.
\end{aligned}
\]
Common accounts remain; $R_K$ is removed and $A_K$ added. Equal list sizes give $|R_K|=|A_K|$. The positive-count change is
\[
  \Delta N_{+}(K)
  =\sum_{v\in A_K}y(v)-\sum_{v\in R_K}y(v).
\]
The downstream order improves the list only when it adds more positives than it removes. The retained accounts cancel, giving $\mathrm{Pr}_{D}@K-\mathrm{Pr}_{U}@K=\Delta N_{+}(K)/K$. At fixed capacity, the rate and count comparisons therefore favor the same order. Across capacities, a smaller rate gap can still mean a larger net positive-count loss as the denominator grows. We track both crossover capacity and the largest pre-crossover positive-count loss. Rate describes queue density; count describes positive starting accounts gained or lost. This decomposition applies to any two capacity-matched account orders, without granting every connected account a review slot.

We increase $K$ from the first operational capacity to the point at which the upstream candidate pool is exhausted. The \emph{initial lead} records which list has the higher positive rate at the first evaluated $K$. A \emph{sustained crossover} is the first later $K$ at which that ordering reverses for five consecutive capacity values. A downstream advantage already present at the first $K$ is an initial lead, not a crossover. At $K=|\mathcal{C}|$, the upstream list contains the full candidate pool, whereas the downstream list may still contain connected accounts from $V\setminus\mathcal{C}$; equality at this endpoint is therefore not forced.

\section{Experimental Design}
\label{sec:experimental-design}

We organize the evaluation around four research questions.
\begin{itemize}
  \item \textbf{RQ1:} Can transaction ranking improve while the positive rate of the final account queue declines?
  \item \textbf{RQ2:} How does the better account order change as the planned review capacity $K$ increases?
  \item \textbf{RQ3:} Which retained, removed, and added accounts explain each change in the queue's positive rate?
  \item \textbf{RQ4:} Can validation-fitted mixing protect a higher-yield order and admit GNN influence as its lead weakens?
\end{itemize}

\subsection{Main Datasets and Candidate Pools}
\label{sec:datasets}

We use two public synthetic AML benchmarks with account-disjoint graph partitions and no cross-partition transactions. Test labels remain hidden until all model and policy choices are frozen.
\begin{itemize}
  \item \textbf{AMLWorld LI-Small}~\cite{altman2023amlworld}: our account projection of the released data contains 705,907 accounts, including 5,304 accounts involved in labeled laundering transactions. Its test partition has 141,389 accounts and 1,213 positives (0.86\%). Endpoints of flagged laundering transactions are positive.
  \item \textbf{AMLNet}~\cite{huda2025amlnet,huda2025amlndata}: our transaction-to-account projection of the released dataset contains 11,000 accounts, including 1,487 positive accounts. Its test partition has 2,176 accounts and 288 positives (13.24\%). We label both endpoints of a flagged laundering transaction positive.
\end{itemize}

All three benchmarks, including HOFINET (Section~\ref{sec:hofinet-transfer}), define positives by transaction-endpoint projection rather than independent AML case confirmation. This can expose positive counterparties among stored neighbors, favoring relational recovery. Five-neighbor sampling does not guarantee exposure, and the builder never consults test labels. Ranking decides which candidates displace upstream accounts. We measure that balance without isolating label projection's contribution to recovery.

A common monitoring gate forms the upstream test pool from transfer size, endpoint velocity, matched turnover, and the number of counterparties. Its rule setting is chosen on validation data and then frozen. After duplicate endpoints are removed, the pool contains 6,057 AMLWorld accounts and 1,926 AMLNet accounts. These pools contain 448 of 1,213 AMLWorld test positives (36.9\%) and 284 of 288 AMLNet test positives (98.6\%), creating complementary low- and high-coverage evaluation surfaces. The AMLWorld validation grid contained no setting within the prespecified alert false-positive-share target (one minus alert precision) of $0.90\pm0.02$; following the frozen selection rule, we used the closest validation setting (0.985). Crossover and recovery are conditional on this frozen surface, which gives both selectors the same starting pool.

\subsection{Account Selectors and GNNs}
\label{sec:models}

\subsubsection{Upstream Selectors}
CatBoost~\cite{prokhorenkova2018catboost} ranks accounts from 12 aggregate features covering transfer volume, direction, counterparty breadth, and coarse timing. Our CLeAR-style selector~\cite{huang2024clear} jointly trains intensity-gated, relative-time sequence encoding with focal classification and two-view contrastive consistency. At each $K$, higher-yield and lower-yield describe the evaluated test orders' $\mathrm{Pr}@K$, not fixed selector identities. Deployment identifies the regime on validation data at the planned capacity.

\subsubsection{Downstream Models}
Multi-GNN, our primary model, operates on directed multigraphs of transfers~\cite{egressy2024directedmultigraph}. GraphSAGE provides an architecture-level confirmation~\cite{hamilton2017sage}. Both models use the same fixed account--transaction graph and the same bounded one-hop surface builder. For every paired run, one frozen GNN account-score vector is reused with both upstream selectors and across all capacities. CLeAR and CatBoost comparisons thus differ in the accounts that open each surface, not in GNN training.

For the primary money-flow measurement, separately fitted logistic decoders rank transactions from endpoint GNN or selector scores. Paired runs use identical four-feature construction, standardization, decoder, candidates, labels, and disjoint validation surface; only the score source differs. Hidden-state and score-induced audits are secondary diagnostics, not selector-specific account results.

\subsection{Training and Repeated Runs}
\label{sec:training}

Selectors and GNNs are trained on the training partition, while validation data select configurations, checkpoints, endpoint aggregation, and score-combination weights. The held-out test partition evaluates only frozen account and relation orders. The GNNs use two width-64 layers, five neighbors per layer, AdamW, focal loss, dropout, and validation-based early stopping.

The capacity sweep uses ten independently fitted runs of each selector, paired one-to-one with ten distinct GNN initializations. Table~\ref{tab:main-fixed-k} reports paired means, sample standard deviations, and paired-bootstrap 95\% confidence intervals (CIs). Exact two-sided sign tests use Holm correction across the eight model-specific account comparisons; pooled architecture summaries form a separate four-comparison family. The primary transaction audit also uses ten distinct selector--GNN pairs. Each pair fits its own dimension-matched GNN-score and selector-score transaction decoders on its own validation surface, and Holm correction is applied across the two GNN comparisons within each dataset. The secondary representation family retains two models and two relation types per dataset. The capacity-specific score-combination study uses the same ten distinct selector--GNN pairs.

\subsection{Capacity Evaluation}
\label{sec:evaluation-by-k}

The primary sweep begins at $K=100$ as an interpretable operational checkpoint. A longitudinal Bank X case study illustrates alert yield using an example monthly review of 100 alerts~\cite{demetis2018fighting}; this motivates the list-size scale, not an observed account budget or a calendar duration. Our experiment varies queue size on fixed test partitions. We then evaluate every integer $K$ from 100 through the upstream pool size so that the findings do not depend on this checkpoint. At each $K$, we take the first $K$ upstream accounts, add their stored one-hop neighbors, and compare $U_K$ with the top-$K$ frozen GNN order restricted to that surface. We record upstream and downstream $\mathrm{Pr}@K$, positive counts, connected accounts, and the retained, removed, and added sets in Section~\ref{sec:review-capacity}. A fixed-max-surface control builds the surface once from the full candidate pool and varies only final-list capacity, isolating it from graph formation.

For a model-specific crossover, the mean downstream positive count must exceed the upstream count in at least eight of ten account-list runs for five consecutive values of $K$. The pooled marker uses the mean over both GNNs and all runs. A downstream lead from $K=100$ is initial; a crossover requires a later reversal.

Score combination compares the upstream order, the GNN order, their equal rank average, and a validation-fitted rank average. We use normalized ranks because raw selector and GNN scores need not be commensurate. Rank combination is invariant to monotone score transformations. Calibration and prior adjustment instead address probabilistic interpretation and class-prior shift~\cite{guo2017calibration,saerens2002prior}; these probabilistic corrections have a different objective from capacity-matched account selection. Only the final account order changes; every condition retains GNN money-flow ranking. We evaluate $K=25$, 50, 100, 150, 500, and 1,000, then extend through the higher-yield-order crossovers at 3,916 on AMLWorld and 1,621 on AMLNet. Let $r_U(v)$ and $r_D(v)$ denote normalized upstream and downstream ranks, with $r_U(v)=0$ for graph-only accounts. An upstream-rank weight of one therefore recovers the upstream queue without removing the GNN. Each test capacity maps to the same validation-pool fraction. For each dataset, selector--GNN pair, run, and capacity, validation selects $\alpha\in\{0,0.1,\ldots,1\}$ to maximize the positive count under $\alpha r_U(v)+(1-\alpha)r_D(v)$, breaking ties toward larger upstream weight. The frozen weight is then evaluated on test data. We denote an alternative order's positive-rate change from upstream by $\Delta\mathrm{Pr}@K$ and report it in percentage points (pp). No account order reserves review positions for connected accounts.

\section{Experimental Results}
\label{sec:results}

\subsection{Money-Flow vs. Account Ranking (RQ1)}
\label{sec:rq1}

Figure~\ref{fig:rq1-disagreement} compares independently learned transaction ranking with account selection on AMLWorld. With both transaction decoders restricted to the same four endpoint features, GNN account scores improve average precision by 0.033 for Multi-GNN and 0.065 for GraphSAGE over CLeAR scores, while account ordering from the same GNN instances removes 6.0 and 7.5 positives from the higher-yield CLeAR list at $K=100$. Both directions hold in all ten paired runs for each architecture, with bootstrap intervals excluding zero and Holm-adjusted exact $p=0.0039$ for transaction ranking and $p=0.0078$ for account ordering. Thus the transaction advantage cannot be attributed to giving the GNN decoder a larger input representation. Table~\ref{tab:main-fixed-k} remains the primary account result from the prespecified capacity sweep; the independent run sets yield close but nonidentical losses. Per-run validation surfaces do not reliably contain both classes of two-transaction paths, so path ranking is excluded from this claim and retained only as a secondary diagnostic. Together, these controls show that better transaction tracing need not imply better selection for limited review slots.

\ifdefined\KoreanPaper
  \def\RelGainY{AP 향상}
  \def\RelGainTitle{\textbf{(a) 동일 차원 거래 AP $\uparrow$}}
  \def\AccountLossTitle{\textbf{(b) 양성 계좌 수 변화 ($K=100$)}}
  \def\RelGainCaption{AMLWorld에서 CLeAR와 $K=100$을 사용한 동일 차원 거래 디코더의 AP 향상과 동일 GNN 실행의 계좌 목록 변화(대응 실행 10회, 수염은 95\% 부트스트랩 구간). AP는 평균정밀도이며, 양쪽 디코더는 동일한 4차원 양끝 계좌 특징을 사용한다.}
  \def\RelGainDescription{GNN 점수와 CLeAR 점수에 동일한 4차원 특징을 사용한 거래 디코더의 평균정밀도는 높아지지만, 같은 GNN의 최종 상위 100개 계좌 목록에서는 양성 계좌 수가 줄어드는 두 개의 막대그래프.}
\else
  \def\RelGainY{AP gain}
  \def\RelGainTitle{\textbf{(a) Dimension-matched Tx AP $\uparrow$}}
  \def\AccountLossTitle{\textbf{(b) Positive@100 $\downarrow$}}
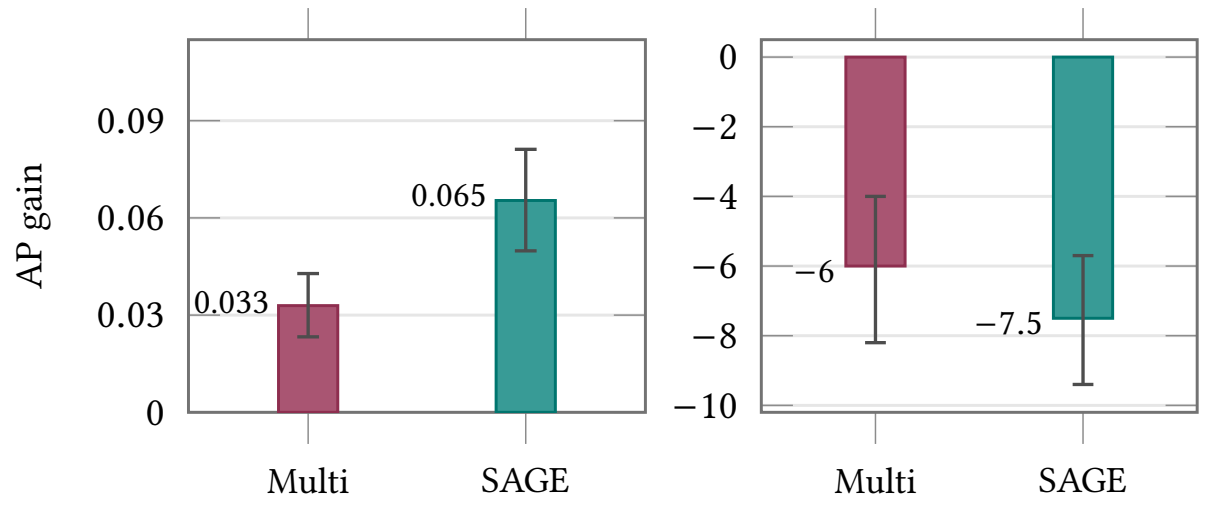
  \def\RelGainCaption{Dimension-matched transaction-decoder AP gain and account-list change on AMLWorld with CLeAR at $K=100$ (ten paired runs; whiskers show 95\% bootstrap intervals). Both decoders use the same four endpoint features.}
  \def\RelGainDescription{Two compact bar charts. Transaction decoders using the same four-dimensional features from GNN or CLeAR account scores improve average precision on the GNN side, while the same GNN instances reduce the number of positive accounts in the final top-100 list.}
\fi
\begin{figure}[t]
\centering
\definecolor{reversalwine}{RGB}{150,48,83}
\definecolor{reversalteal}{RGB}{0,127,120}
\resizebox{.94\columnwidth}{!}{%
\begin{tikzpicture}[font=\sffamily]
  \begin{axis}[
    name=relationgain,
    width=3.75cm,height=3.35cm,
    xmin=.45,xmax=2.55,ymin=0,ymax=.115,
    xtick={1,2},xticklabels={Multi,SAGE},
    ytick={0,.03,.06,.09},
    yticklabel style={/pgf/number format/fixed,/pgf/number format/precision=2},
    ylabel={\RelGainY},
    tick label style={font=\fontsize{5.0}{5.5}\selectfont},
    label style={font=\fontsize{5.2}{5.7}\selectfont},
    axis line style={black!55},tick style={black!45},
    ymajorgrids,grid style={black!10},
    ybar=0pt,bar width=8.0pt,
  ]
    \addplot[fill=reversalwine!82,draw=reversalwine!95!black,bar shift=0pt]
      coordinates {(1,.032931)};
    \addplot[fill=reversalteal!78,draw=reversalteal!92!black,bar shift=0pt]
      coordinates {(2,.065373)};
    \draw[black!70,line width=.45pt] (axis cs:1,.023295)--(axis cs:1,.042824);
    \draw[black!70,line width=.45pt] (axis cs:.95,.023295)--(axis cs:1.05,.023295);
    \draw[black!70,line width=.45pt] (axis cs:.95,.042824)--(axis cs:1.05,.042824);
    \draw[black!70,line width=.45pt] (axis cs:2,.049829)--(axis cs:2,.081160);
    \draw[black!70,line width=.45pt] (axis cs:1.95,.049829)--(axis cs:2.05,.049829);
    \draw[black!70,line width=.45pt] (axis cs:1.95,.081160)--(axis cs:2.05,.081160);
    \node[font=\fontsize{4.9}{5.3}\selectfont,anchor=east]
      at (axis cs:.93,.034) {0.033};
    \node[font=\fontsize{4.9}{5.3}\selectfont,anchor=east]
      at (axis cs:1.93,.067) {0.065};
  \end{axis}

  \begin{axis}[
    name=accountloss,
    at={($(relationgain.south east)+(0.55cm,0)$)},anchor=south west,
    width=3.65cm,height=3.35cm,
    xmin=.45,xmax=2.55,ymin=-10.2,ymax=.5,
    xtick={1,2},xticklabels={Multi,SAGE},
    ytick={-10,-8,-6,-4,-2,0},
    tick label style={font=\fontsize{5.0}{5.5}\selectfont},
    label style={font=\fontsize{5.2}{5.7}\selectfont},
    axis line style={black!55},tick style={black!45},
    ymajorgrids,grid style={black!10},
    ybar=0pt,bar width=8.0pt,
  ]
    \addplot[fill=reversalwine!82,draw=reversalwine!95!black,bar shift=0pt]
      coordinates {(1,-6.0)};
    \addplot[fill=reversalteal!78,draw=reversalteal!92!black,bar shift=0pt]
      coordinates {(2,-7.5)};
    \draw[black!70,line width=.45pt] (axis cs:1,-8.2)--(axis cs:1,-4.0);
    \draw[black!70,line width=.45pt] (axis cs:.95,-8.2)--(axis cs:1.05,-8.2);
    \draw[black!70,line width=.45pt] (axis cs:.95,-4.0)--(axis cs:1.05,-4.0);
    \draw[black!70,line width=.45pt] (axis cs:2,-9.4)--(axis cs:2,-5.7);
    \draw[black!70,line width=.45pt] (axis cs:1.95,-9.4)--(axis cs:2.05,-9.4);
    \draw[black!70,line width=.45pt] (axis cs:1.95,-5.7)--(axis cs:2.05,-5.7);
    \node[font=\fontsize{4.9}{5.3}\selectfont,anchor=east]
      at (axis cs:.92,-6.15) {$-6$};
    \node[font=\fontsize{4.9}{5.3}\selectfont,anchor=east]
      at (axis cs:1.92,-7.65) {$-7.5$};
  \end{axis}
  \node[font=\bfseries\fontsize{5.5}{6.0}\selectfont,anchor=north]
    at ($(relationgain.south)+(0,-.50cm)$) {\RelGainTitle};
  \node[font=\bfseries\fontsize{5.5}{6.0}\selectfont,anchor=north]
    at ($(accountloss.south)+(0,-.50cm)$) {\AccountLossTitle};
\end{tikzpicture}%
}
\caption{\RelGainCaption}
\Description{\RelGainDescription}
\label{fig:rq1-disagreement}
\end{figure}

\begingroup
\begin{table}[h]
  \centering
  \caption{Fixed-$K$ account results at $K=100$ from ten independent selector--GNN pairs. Positive change reports mean $\pm$ sample standard deviation and a paired bootstrap interval.}
  \label{tab:main-fixed-k}
  \small
  \setlength{\tabcolsep}{0.5pt}
  \ifdefined\KoreanPaper
  \def\MainSetting{조건}
  \def\MainDeltaHead{양성 변화}
  \def\MainMeanSd{평균 $\pm$ 표준편차}
  \def\MainCI{95\% 구간}
  \def\MainUp{선별}
  \def\MainDown{GNN 순위}
\else
  \def\MainSetting{Setting}
  \def\MainDeltaHead{Positive $\Delta$}
  \def\MainMeanSd{Mean $\pm$ SD}
  \def\MainCI{95\% CI}
  \def\MainUp{Upstream}
  \def\MainDown{Downstream}
\fi
\begin{tabular}{@{}cccccc@{}}
\toprule
\multirow{2}{*}{\MainSetting} & \multirow{2}{*}{GNN} & \multicolumn{2}{c}{Pr@100} & \multicolumn{2}{c}{\MainDeltaHead} \\
\cmidrule(lr){3-4}\cmidrule(l){5-6}
 & & \MainUp & \MainDown & \MainMeanSd & \MainCI \\
\midrule
\multirow{2}{*}{\makecell[l]{AMLWorld\\CLeAR}} & Multi-GNN & \multirow{2}{*}{0.89} & 0.83 & $-\phantom{0}5.7_{\pm \phantom{0}5.9}$ & $[-\phantom{0}9.2,-\phantom{0}2.2]$ \\
 & GraphSAGE & & 0.82 & $-\phantom{0}7.1_{\pm \phantom{0}4.2}$ & $[-\phantom{0}9.5,-\phantom{0}4.6]$ \\
\cmidrule(lr){1-6}
\multirow{2}{*}{\makecell[l]{AMLWorld\\CatBoost}} & Multi-GNN & \multirow{2}{*}{0.39} & 0.42 & $+\phantom{0}3.0_{\pm \phantom{0}3.4}$ & $[+\phantom{0}0.9,+\phantom{0}4.9]$ \\
 & GraphSAGE & & 0.44 & $+\phantom{0}4.7_{\pm \phantom{0}3.7}$ & $[+\phantom{0}2.7,+\phantom{0}7.1]$ \\
\midrule
\multirow{2}{*}{\makecell[l]{AMLNet\\CLeAR}} & Multi-GNN & \multirow{2}{*}{0.68} & 0.87 & $+19.8_{\pm 19.4}$ & $[+\phantom{0}8.8,+31.4]$ \\
 & GraphSAGE & & 0.86 & $+18.0_{\pm 19.4}$ & $[+\phantom{0}6.6,+29.3]$ \\
\cmidrule(lr){1-6}
\multirow{2}{*}{\makecell[l]{AMLNet\\CatBoost}} & Multi-GNN & \multirow{2}{*}{0.97} & 0.96 & $-\phantom{0}1.4_{\pm \phantom{0}2.2}$ & $[-\phantom{0}2.9,-\phantom{0}0.3]$ \\
 & GraphSAGE & & 0.95 & $-\phantom{0}2.7_{\pm \phantom{0}5.6}$ & $[-\phantom{0}6.4,-\phantom{0}0.3]$ \\
\bottomrule
\end{tabular}

\end{table}

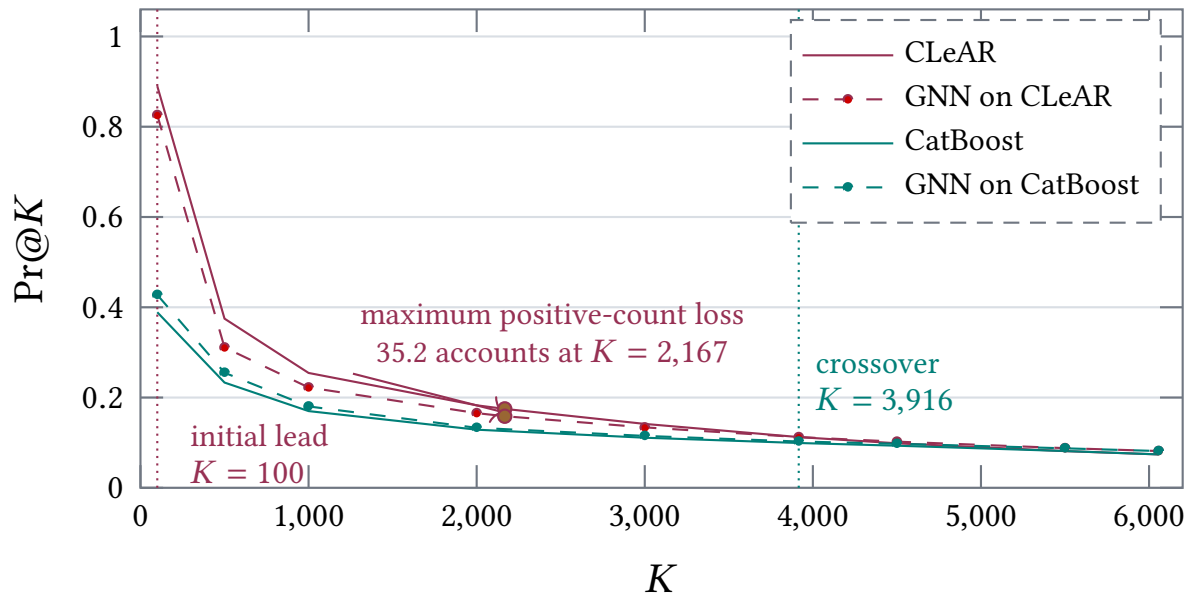
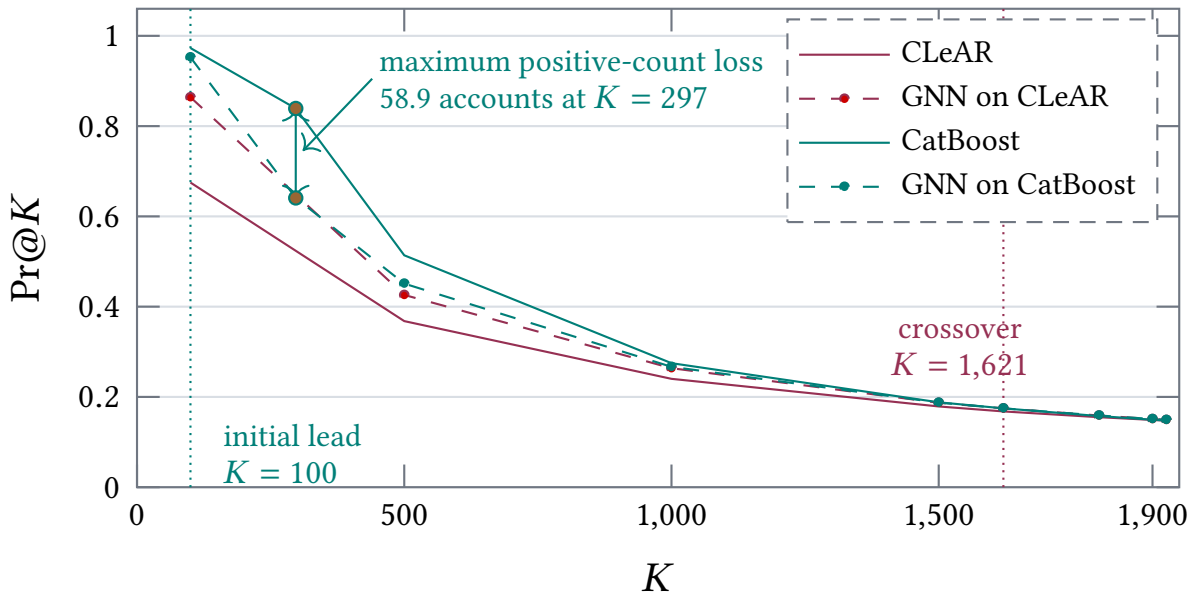
\begin{figure}[!t]
  \centering
  \definecolor{doitwine}{HTML}{963053}
\definecolor{doitteal}{HTML}{007F78}
\definecolor{doitgray}{HTML}{707782}
\definecolor{doitgrid}{HTML}{D9DEE4}
\ifdefined\KoreanPaper
  \def\CapacityYLabel{양성률}
  \def\CapacityInitial{초기 우위}
  \def\CapacityCross{역전}
  \def\CapacityPeakLoss{최대 양성 계좌 손실}
  \def\CapacityAccounts{개}
  \def\CapacityAt{,}
  \def\CapacityCLeARGNN{CLeAR 표면의 GNN}
  \def\CapacityCatGNN{CatBoost 표면의 GNN}
\else
  \def\CapacityYLabel{$\mathrm{Pr}@K$}
  \def\CapacityInitial{initial lead}
  \def\CapacityCross{crossover}
  \def\CapacityPeakLoss{maximum positive-count loss}
  \def\CapacityAccounts{accounts}
  \def\CapacityAt{ at}
  \def\CapacityCLeARGNN{GNN on CLeAR}
  \def\CapacityCatGNN{GNN on CatBoost}
\fi
\pgfplotsset{
doitcapacity/.style={
  width=\linewidth,
  height=0.56\linewidth,
  ymin=0,
  ymax=1.06,
  ytick={0,0.2,0.4,0.6,0.8,1.0},
  axis line style={draw=doitgray,thin},
  tick style={draw=doitgray,thin},
  tick label style={font=\scriptsize,text=black},
  label style={font=\small,text=black},
  title style={font=\small\bfseries,text=black},
  grid=major,
  xmajorgrids=false,
  grid style={draw=doitgrid,thin},
  legend style={
    font=\scriptsize,
    at={(0.98,0.98)},
    anchor=north east,
    draw=doitgray,
    densely dashed,
    fill=white,
    inner sep=2pt,
    row sep=-1pt
  },
  legend cell align={left},
  clip=false
}}

\begin{subfigure}[t]{\linewidth}
\centering
\begin{tikzpicture}
\begin{axis}[
  doitcapacity,
  xlabel={$K$},
  ylabel={\CapacityYLabel},
  xmin=0,
  xmax=6200,
  xtick={0,1000,2000,3000,4000,5000,6000}
]
\addplot+[color=doitwine,solid,thin,mark=none] coordinates {(100,0.8900) (500,0.3748) (1000,0.2545) (2000,0.1828) (2167,0.1749) (3000,0.1416) (3916,0.1124) (4500,0.0986) (5500,0.0813) (6057,0.0740)};
\addlegendentry{CLeAR}
\addplot+[color=doitwine,dashed,thin,mark=*,mark size=0.8pt] coordinates {(100,0.8260) (500,0.3110) (1000,0.2224) (2000,0.1655) (2167,0.1586) (3000,0.1339) (3916,0.1124) (4500,0.1023) (5500,0.0879) (6057,0.0815)};
\addlegendentry{\CapacityCLeARGNN}
\addplot+[color=doitteal,solid,thin,mark=none] coordinates {(100,0.3890) (500,0.2332) (1000,0.1700) (2000,0.1288) (3000,0.1110) (3916,0.0993) (4500,0.0933) (5500,0.0815) (6057,0.0740)};
\addlegendentry{CatBoost}
\addplot+[color=doitteal,dashed,thin,mark=*,mark size=0.8pt] coordinates {(100,0.4275) (500,0.2550) (1000,0.1803) (2000,0.1330) (3000,0.1150) (3916,0.1026) (4500,0.0976) (5500,0.0875) (6057,0.0815)};
\addlegendentry{\CapacityCatGNN}
\addplot+[color=doitwine,densely dotted,thin,mark=none] coordinates {(100,0) (100,1.06)};
  \node[font=\scriptsize,text=doitwine,align=left,inner sep=1pt,anchor=west] at (axis cs:260,0.075) {\CapacityInitial\\$K=100$};
\addplot+[color=doitteal,densely dotted,thin,mark=none] coordinates {(3916,0) (3916,1.06)};
  \node[font=\scriptsize,text=doitteal,align=left,inner sep=1pt,anchor=west] at (axis cs:3980,0.225) {\CapacityCross\\$K=3{,}916$};
\addplot+[color=doitwine,only marks,mark=*,mark size=1.25pt,forget plot]
  coordinates {(2167,0.1749) (2167,0.1586)};
\node[font=\scriptsize,text=doitwine,align=center,inner sep=1.2pt]
  (awpeak) at (axis cs:2450,0.340) {\CapacityPeakLoss\\35.2 \CapacityAccounts\CapacityAt{} $K=2{,}167$};
\draw[->,color=doitwine,thin] (awpeak.south west) -- (axis cs:2167,0.1667);
\end{axis}
\end{tikzpicture}
\caption{AMLWorld}
\label{fig:capacity-crossover-amlworld}
\end{subfigure}

\vspace{2pt}

\begin{subfigure}[t]{\linewidth}
\centering
\begin{tikzpicture}
\begin{axis}[
  doitcapacity,
  xlabel={$K$},
  ylabel={\CapacityYLabel},
  xmin=0,
  xmax=1950,
  xtick={0,500,1000,1500,1900}
]
\addplot+[color=doitwine,solid,thin,mark=none] coordinates {(100,0.6750) (500,0.3680) (1000,0.2401) (1500,0.1789) (1621,0.1679) (1800,0.1551) (1900,0.1491) (1926,0.1475)};
\addlegendentry{CLeAR}
\addplot+[color=doitwine,dashed,thin,mark=*,mark size=0.8pt] coordinates {(100,0.8640) (500,0.4260) (1000,0.2637) (1500,0.1871) (1621,0.1747) (1800,0.1588) (1900,0.1510) (1926,0.1491)};
\addlegendentry{\CapacityCLeARGNN}
\addplot+[color=doitteal,solid,thin,mark=none] coordinates {(100,0.9730) (297,0.8391) (500,0.5138) (1000,0.2748) (1500,0.1881) (1621,0.1747) (1800,0.1576) (1900,0.1495) (1926,0.1475)};
\addlegendentry{CatBoost}
\addplot+[color=doitteal,dashed,thin,mark=*,mark size=0.8pt] coordinates {(100,0.9525) (297,0.6409) (500,0.4511) (1000,0.2667) (1500,0.1874) (1621,0.1747) (1800,0.1589) (1900,0.1511) (1926,0.1491)};
\addlegendentry{\CapacityCatGNN}
\addplot+[color=doitteal,densely dotted,thin,mark=none] coordinates {(100,0) (100,1.06)};
  \node[font=\scriptsize,text=doitteal,align=left,inner sep=1pt,anchor=west] at (axis cs:150,0.075) {\CapacityInitial\\$K=100$};
\addplot+[color=doitwine,densely dotted,thin,mark=none] coordinates {(1621,0) (1621,1.06)};
  \node[font=\scriptsize,text=doitwine,align=center,inner sep=1pt] at (axis cs:1540,0.300) {\CapacityCross\\$K=1{,}621$};
\addplot+[color=doitteal,only marks,mark=*,mark size=1.25pt,forget plot]
  coordinates {(297,0.8391) (297,0.6409)};
\draw[<->,color=doitteal,thin] (axis cs:297,0.6409) -- (axis cs:297,0.8391);
\node[font=\scriptsize,text=doitteal,align=left,inner sep=1.2pt,anchor=west]
  (anpeak) at (axis cs:440,0.905) {\CapacityPeakLoss\\58.9 \CapacityAccounts\CapacityAt{} $K=297$};
\draw[->,color=doitteal,thin] (anpeak.west) -- (axis cs:310,0.745);
\end{axis}
\end{tikzpicture}
\caption{AMLNet}
\label{fig:capacity-crossover-amlnet}
\end{subfigure}
  \caption{Positive rate over investigation capacity $K$. Lines average ten account-list runs per GNN and then equally average the two GNNs. Solid and dashed lines denote upstream selector and raw GNN orders on each selector-opened surface; annotations mark initial leads, the largest mean positive-count loss for each higher-yield order, and sustained crossovers.}
  \Description{Two line charts, one for AMLWorld and one for AMLNet, comparing upstream and downstream positive rate as K increases for CLeAR and CatBoost. Callouts mark the largest loss in positive accounts before each higher-yield order reaches its crossover.}
  \label{fig:capacity-crossover}
\end{figure}

Table~\ref{tab:main-fixed-k} shows that this is not a general failure of GNN account scores. Paired bootstrap intervals preserve the direction of all eight account-list changes: GNN ordering improves the lower-yield upstream order in both datasets but underperforms the higher-yield order. Holm-corrected sign tests confirm both directions for AMLWorld GraphSAGE and the AMLNet improvement for Multi-GNN; the remaining model-specific tests are not significant after correction. Bootstrap intervals estimate the mean paired change using effect magnitudes, whereas exact sign tests use only non-tied directions. With ten pairs and Holm correction, the sign test is more conservative; we report both as complementary evidence of mean effect and cross-run consistency. The same downstream model can repair a lower-yield order but damage a higher-yield one. A connected account may clarify a money flow without deserving earlier direct review; account yield and money-flow ranking should therefore be evaluated and reported separately.

Figure~\ref{fig:capacity-crossover} tracks both account orders as investigation capacity grows. A higher-yield upstream order leads initially, whereas a GNN paired with a lower-yield order starts with an \emph{initial lead}. As $K$ grows, the higher-yield order consumes its positive-dense prefix while the graph continues to expose connected accounts.

AMLNet shows the converse separation: its lower-yield account order improves even though the dimension-matched transaction decoders show no stable transaction advantage. These results show that independently decoded transaction ranking and account selection can move in opposite directions; transaction improvement is neither necessary nor sufficient for higher account yield.

\subsection{Account Ordering by Capacity (RQ2)}
\label{sec:rq2}

Both pooled reversals occur only after most of the upstream pool is under review. The pooled crossovers are $K=3{,}916$ on AMLWorld and 1,621 on AMLNet; Multi-GNN and GraphSAGE cross at 3,947 and 4,473 on AMLWorld and at 1,543 and 1,842 on AMLNet. Because pooling uses architecture means while model-specific crossings also require run-level consistency, the pooled marker need not fall between them. The fixed $K=100$ checkpoint understates the worst displacement: equally averaging both GNNs, AMLWorld CLeAR grows from 6.4 fewer positives at $K=100$ to a first empirical maximum of 35.2 at $K=2{,}167$, while AMLNet CatBoost grows from 2.1 to 58.9 at $K=297$. These surface-specific peaks identify maximum pre-crossover losses, not recommended quotas.

This cost matters because account selection determines where downstream graph investigation begins; even better transaction ranking cannot compensate when its account order replaces a higher-yield starting set at the same capacity. A validated higher-yield order should therefore open the review surface, with the GNN ranking transaction evidence there and altering the queue only when capacity-matched validation supports it. Crossover values remain dataset-, selector-, graph-, and capacity-range-specific.

The fixed-max-surface control preserves this late reversal for the higher-yield orders, establishing that the pattern survives when graph formation is held constant. Its pooled crossover moves from $K=3{,}916$ to 4,632 on AMLWorld and from $K=1{,}621$ to 1,645 on AMLNet. Graph formation is therefore an operational policy variable: AMLWorld CatBoost has an operational GNN initial lead, whereas its fixed-surface order crosses at $K=3{,}082$. Candidate-pool coverage provides a second boundary condition: AMLWorld's 36.9\% coverage leaves positives outside the upstream pool for graph expansion to expose, AMLNet's 98.6\% still crosses, and HOFINET's complete pool shows no sustained crossover. Coverage thus bounds external positive headroom but does not by itself determine whether or where a crossover occurs. Because the regime depends on quantities an institution can measure, it can be determined on held-out historical cases at the planned $K$: validation $\mathrm{Pr}@K$ identifies the higher-yield upstream order, and the same validation surface decides whether GNN reranking improves it and selects the final rank weight. Table~\ref{tab:capacity-guide} summarizes the observed account-order regimes.

\begin{table}[h]
  \centering
  \captionsetup{justification=raggedright,singlelinecheck=false}
  \caption{Capacity-matched account-order guide. The GNN ranks money-flow evidence in every regime.}
  \label{tab:capacity-guide}
  \small
  \setlength{\tabcolsep}{2.0pt}



\ifdefined\KoreanPaper
  \def\GuideRegime{관측 조건}
  \def\GuideOrder{계좌 순위 규칙}
  \def\GuideControl{GNN 계좌 점수 통제}

  \def\GuideHigh{고수율 순위,\newline 작은 $K$}
  \def\GuideHighOrder{선별 또는\newline 검증 결합}
  \def\GuideHighControl{선별 우위\newline 보호}

  \def\GuideLow{저수율 순위,\newline 작은 $K$}
  \def\GuideLowOrder{검증 데이터가\newline 선택한 순위}
  \def\GuideLowControl{검증될 때만\newline 영향 허용}

  \def\GuideCross{역전 부근\newline 및 이후}
  \def\GuideCrossOrder{검토 건수별\newline 검증}
  \def\GuideCrossControl{향상이 유지될 때만\newline 영향 확대}
\else
  \def\GuideRegime{Regime}
  \def\GuideOrder{Account-order rule}
  \def\GuideControl{GNN account-score control}

  \def\GuideHigh{Higher-yield,\newline small $K$}
  \def\GuideHighOrder{Upstream or\newline validated mix}
  \def\GuideHighControl{Protect upstream\newline lead}

  \def\GuideLow{Lower-yield,\newline small $K$}
  \def\GuideLowOrder{Validation-selected\newline order}
  \def\GuideLowControl{Admit only\newline if validated}

  \def\GuideCross{Near or beyond\newline crossover}
  \def\GuideCrossOrder{Capacity-matched\newline validation}
  \def\GuideCrossControl{Increase only while\newline gain persists}
\fi

\begingroup
\renewcommand{\arraystretch}{1.15}
\setlength{\tabcolsep}{3pt}

\renewcommand{\tabularxcolumn}[1]{m{#1}}

\begin{tabularx}{\columnwidth}{
  @{}
  >{\centering\arraybackslash}m{0.24\columnwidth}
  >{\centering\arraybackslash}m{0.31\columnwidth}
  >{\centering\arraybackslash}X
  @{}
}
\toprule
\GuideRegime & \GuideOrder & \GuideControl \\
\midrule
\GuideHigh & \GuideHighOrder & \GuideHighControl \\
\addlinespace[5pt]
\GuideLow & \GuideLowOrder & \GuideLowControl \\
\addlinespace[5pt]
\GuideCross & \GuideCrossOrder & \GuideCrossControl \\
\bottomrule
\end{tabularx}

\endgroup
\end{table}

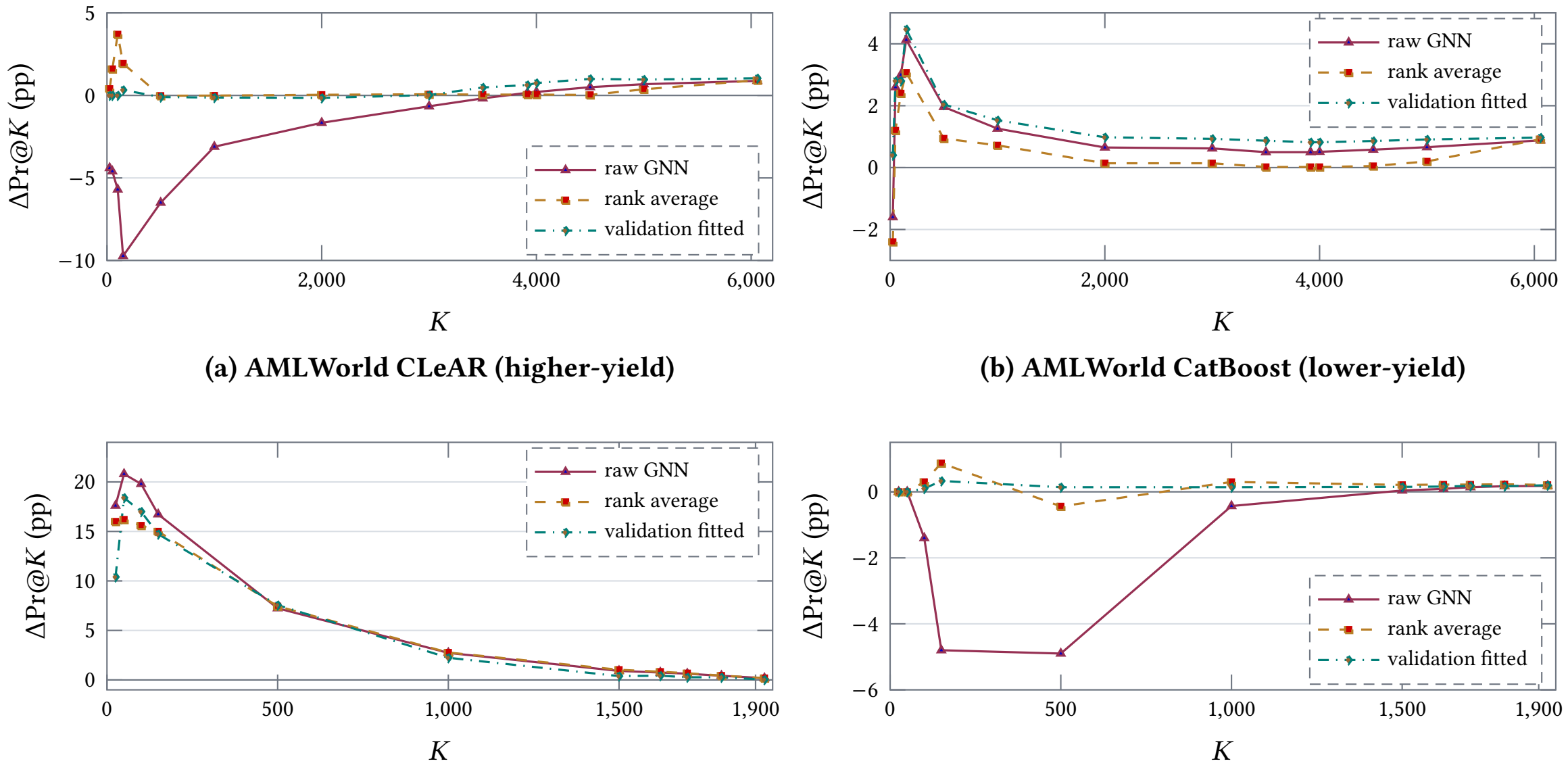
\begin{figure*}[t]
  \centering
  \definecolor{doitwine}{HTML}{963053}
\definecolor{doitteal}{HTML}{007F78}
\definecolor{doitorange}{HTML}{B87E26}
\definecolor{doitblack}{HTML}{24292F}
\definecolor{doitgray}{HTML}{707782}
\definecolor{doitgrid}{HTML}{D9DEE4}
\ifdefined\KoreanPaper
  \def\ScoreTitleAWC{AMLWorld CLeAR (고수율)}
  \def\ScoreTitleAWB{AMLWorld CatBoost (저수율)}
  \def\ScoreTitleANC{AMLNet CLeAR (저수율)}
  \def\ScoreTitleANB{AMLNet CatBoost (고수율)}
  \def\ScoreYLabel{$\Delta$양성률 (\%p)}
  \def\ScoreRaw{GNN 원순위}
  \def\ScoreAverage{순위 평균}
  \def\ScoreFitted{검증 최적}
  \def\ScorePanelBYMin{-2}
\else
  \def\ScoreTitleAWC{AMLWorld CLeAR (higher-yield)}
  \def\ScoreTitleAWB{AMLWorld CatBoost (lower-yield)}
  \def\ScoreTitleANC{AMLNet CLeAR (lower-yield)}
  \def\ScoreTitleANB{AMLNet CatBoost (higher-yield)}
  \def\ScoreYLabel{$\Delta\mathrm{Pr}@K$ (pp)}
  \def\ScoreRaw{raw GNN}
  \def\ScoreAverage{rank average}
  \def\ScoreFitted{validation fitted}
  \def\ScorePanelBYMin{-3}
\fi
\begin{tikzpicture}
\begin{groupplot}[
  group style={group size=2 by 2,horizontal sep=1.15cm,vertical sep=1.78cm},
  width=0.455\textwidth,
  height=0.225\textwidth,
  y filter/.expression={y*100},
  axis line style={draw=doitgray,thin},
  tick style={draw=doitgray,thin},
  tick label style={font=\scriptsize,text=black},
  label style={font=\small,text=black},
  grid=major,
  xmajorgrids=false,
  grid style={draw=doitgrid,thin},
  legend style={
    font=\scriptsize,
    at={(0.98,0.98)},
    anchor=north east,
    draw=doitgray,
    densely dashed,
    fill=white,
    inner sep=2pt,
    row sep=-1pt
  },
  legend cell align={left},
  clip=false
]

\nextgroupplot[
  xlabel={$K$},ylabel={\ScoreYLabel},
  xmin=0,xmax=6200,xtick={0,2000,4000,6000},
  ymin=-10,ymax=5,ytick={-10,-5,0,5},
  legend style={at={(0.98,0.02)},anchor=south east,legend columns=1}
]
\addplot+[color=doitgray,solid,thin,mark=none,forget plot] coordinates {(0,0) (6200,0)};
\addplot+[color=doitwine,solid,semithick,mark=triangle*,mark size=1.15pt] coordinates {(25,-0.0440) (50,-0.0460) (100,-0.0570) (150,-0.0973) (500,-0.0650) (1000,-0.0311) (2000,-0.0166) (3000,-0.0066) (3500,-0.0018) (3916,0.0016) (4000,0.0021) (4500,0.0050) (5000,0.0068) (6057,0.0088)};
\addlegendentry{\ScoreRaw}
\addplot+[color=doitorange,dashed,semithick,mark=square*,mark size=1.0pt] coordinates {(25,0.0040) (50,0.0160) (100,0.0370) (150,0.0193) (500,-0.0002) (1000,-0.0001) (2000,0.0004) (3000,0.0007) (3500,0.0005) (3916,0.0005) (4000,0.0005) (4500,0.0003) (5000,0.0037) (6057,0.0092)};
\addlegendentry{\ScoreAverage}
\addplot+[color=doitteal,dash dot,semithick,mark=diamond*,mark size=1.15pt] coordinates {(25,0.0000) (50,0.0000) (100,0.0000) (150,0.0033) (500,-0.0010) (1000,-0.0013) (2000,-0.0015) (3000,0.0002) (3500,0.0048) (3916,0.0063) (4000,0.0075) (4500,0.0100) (5000,0.0096) (6057,0.0104)};
\addlegendentry{\ScoreFitted}

\nextgroupplot[
  xlabel={$K$},ylabel={\ScoreYLabel},
  xmin=0,xmax=6200,xtick={0,2000,4000,6000},
  ymin=\ScorePanelBYMin,ymax=5,ytick={-2,0,2,4}
]
\addplot+[color=doitgray,solid,thin,mark=none,forget plot] coordinates {(0,0) (6200,0)};
\addplot+[color=doitwine,solid,semithick,mark=triangle*,mark size=1.15pt] coordinates {(25,-0.0160) (50,0.0260) (100,0.0300) (150,0.0413) (500,0.0196) (1000,0.0126) (2000,0.0065) (3000,0.0062) (3500,0.0050) (3916,0.0050) (4000,0.0051) (4500,0.0058) (5000,0.0066) (6057,0.0088)};
\addlegendentry{\ScoreRaw}
\addplot+[color=doitorange,dashed,semithick,mark=square*,mark size=1.0pt] coordinates {(25,-0.0240) (50,0.0120) (100,0.0240) (150,0.0307) (500,0.0094) (1000,0.0072) (2000,0.0014) (3000,0.0014) (3500,0.0002) (3916,0.0002) (4000,0.0002) (4500,0.0005) (5000,0.0020) (6057,0.0092)};
\addlegendentry{\ScoreAverage}
\addplot+[color=doitteal,dash dot,semithick,mark=diamond*,mark size=1.15pt] coordinates {(25,0.0040) (50,0.0280) (100,0.0280) (150,0.0447) (500,0.0204) (1000,0.0153) (2000,0.0098) (3000,0.0093) (3500,0.0087) (3916,0.0082) (4000,0.0082) (4500,0.0086) (5000,0.0091) (6057,0.0097)};
\addlegendentry{\ScoreFitted}

\nextgroupplot[
  xlabel={$K$},ylabel={\ScoreYLabel},
  xmin=0,xmax=1950,xtick={0,500,1000,1500,1900},
  ymin=-1,ymax=24,ytick={0,5,10,15,20}
]
\addplot+[color=doitgray,solid,thin,mark=none,forget plot] coordinates {(0,0) (1950,0)};
\addplot+[color=doitwine,solid,semithick,mark=triangle*,mark size=1.15pt] coordinates {(25,0.1760) (50,0.2080) (100,0.1980) (150,0.1673) (500,0.0724) (1000,0.0273) (1500,0.0091) (1621,0.0075) (1700,0.0063) (1800,0.0042) (1926,0.0018)};
\addlegendentry{\ScoreRaw}
\addplot+[color=doitorange,dashed,semithick,mark=square*,mark size=1.0pt] coordinates {(25,0.1600) (50,0.1620) (100,0.1560) (150,0.1500) (500,0.0744) (1000,0.0277) (1500,0.0103) (1621,0.0083) (1700,0.0068) (1800,0.0044) (1926,0.0019)};
\addlegendentry{\ScoreAverage}
\addplot+[color=doitteal,dash dot,semithick,mark=diamond*,mark size=1.15pt] coordinates {(25,0.1040) (50,0.1840) (100,0.1700) (150,0.1473) (500,0.0756) (1000,0.0224) (1500,0.0039) (1621,0.0043) (1700,0.0027) (1800,0.0027) (1926,0.0003)};
\addlegendentry{\ScoreFitted}

\nextgroupplot[
  xlabel={$K$},ylabel={\ScoreYLabel},
  xmin=0,xmax=1950,xtick={0,500,1000,1500,1900},
  ymin=-6,ymax=1.5,ytick={-6,-4,-2,0},
  legend style={at={(0.98,0.02)},anchor=south east,legend columns=1}
]
\addplot+[color=doitgray,solid,thin,mark=none,forget plot] coordinates {(0,0) (1950,0)};
\addplot+[color=doitwine,solid,semithick,mark=triangle*,mark size=1.15pt] coordinates {(25,0.0000) (50,0.0000) (100,-0.0140) (150,-0.0480) (500,-0.0490) (1000,-0.0043) (1500,0.0004) (1621,0.0009) (1700,0.0014) (1800,0.0017) (1926,0.0018)};
\addlegendentry{\ScoreRaw}
\addplot+[color=doitorange,dashed,semithick,mark=square*,mark size=1.0pt] coordinates {(25,0.0000) (50,0.0000) (100,0.0030) (150,0.0087) (500,-0.0044) (1000,0.0030) (1500,0.0021) (1621,0.0022) (1700,0.0022) (1800,0.0023) (1926,0.0021)};
\addlegendentry{\ScoreAverage}
\addplot+[color=doitteal,dash dot,semithick,mark=diamond*,mark size=1.15pt] coordinates {(25,0.0000) (50,0.0000) (100,0.0010) (150,0.0033) (500,0.0014) (1000,0.0014) (1500,0.0015) (1621,0.0016) (1700,0.0018) (1800,0.0017) (1926,0.0020)};
\addlegendentry{\ScoreFitted}
\end{groupplot}
\node[font=\small\bfseries,anchor=north]
  at ($(group c1r1.south)+(0,-.82cm)$) {(a) \ScoreTitleAWC};
\node[font=\small\bfseries,anchor=north]
  at ($(group c2r1.south)+(0,-.82cm)$) {(b) \ScoreTitleAWB};
\node[font=\small\bfseries,anchor=north]
  at ($(group c1r2.south)+(0,-.82cm)$) {(c) \ScoreTitleANC};
\node[font=\small\bfseries,anchor=north]
  at ($(group c2r2.south)+(0,-.82cm)$) {(d) \ScoreTitleANB};
\end{tikzpicture}
  \caption{Positive-rate change relative to the upstream order over investigation capacity $K$ (Multi-GNN; ten independent paired runs). Zero denotes the upstream result; ``validation fitted'' uses a rank weight selected at the matching validation-pool fraction.}
  \Description{Four small difference charts compare raw GNN, rank-average, and validation-fitted account orders with the upstream baseline across K. Values above zero improve upon upstream account selection, and values below zero reduce its positive rate.}
  \label{fig:score-combination}
\end{figure*}

\subsection{Account Replacement Mechanism (RQ3)}
\label{sec:rq3}

Table~\ref{tab:replacement-summary} decomposes each change into positive neighbors added and positive upstream accounts displaced. Because both lists have the same size, every addition has a matched removal, so gains and losses reduce to one balance. The mechanism is the same in both regimes; what differs is the price. Against a higher-yield order, additions displace a positive-dense prefix and do not cover their cost; against a lower-yield order, the same additions replace a sparser set and the sign reverses. Reporting positive neighbors without the accounts they displace would therefore overstate downstream value.

\begin{table}[t]
  \centering
  \caption{Account-set decomposition at $K=100$ (ten-run means). Retained and removed count upstream positives; added counts positive neighbors.}
  \label{tab:replacement-summary}
  \small
  \setlength{\tabcolsep}{1.8pt}
  \ifdefined\KoreanPaper
  \def\ReplaceSetting{조건}
  \def\ReplacePositive{양성 계좌}
  \def\ReplaceNet{\makecell{순변화}}
  \def\ReplaceKept{유지}
  \def\ReplaceAdded{추가}
  \def\ReplaceRemoved{제거}
\else
  \def\ReplaceSetting{Setting}
  \def\ReplacePositive{Positive Accounts}
  \def\ReplaceNet{\makecell{Net\\Change}}
  \def\ReplaceKept{Retained}
  \def\ReplaceAdded{Added}
  \def\ReplaceRemoved{Removed}
\fi
\begin{tabular}{@{}cccccc@{}}
\toprule
\multirow{2}{*}{\ReplaceSetting} & \multirow{2}{*}{GNN} & \multicolumn{3}{c}{\ReplacePositive} & \multirow{2}{*}{\ReplaceNet} \\
\cmidrule(lr){3-5}
 & & \ReplaceKept & \ReplaceAdded & \ReplaceRemoved & \\
\midrule
\multirow{2}{*}{\makecell[l]{AMLWorld\\CLeAR}} & Multi-GNN & 52.7 & 30.6 & 36.3 & $-\phantom{0}5.7$ \\
 & GraphSAGE & 49.9 & 32.0 & 39.1 & $-\phantom{0}7.1$ \\
\cmidrule(lr){1-6}
\multirow{2}{*}{\makecell[l]{AMLWorld\\CatBoost}} & Multi-GNN & 27.2 & 14.7 & 11.7 & $+\phantom{0}3.0$ \\
 & GraphSAGE & 26.5 & 17.1 & 12.4 & $+\phantom{0}4.7$ \\
\midrule
\multirow{2}{*}{\makecell[l]{AMLNet\\CLeAR}} & Multi-GNN & 46.6 & 40.7 & 20.9 & $+19.8$ \\
 & GraphSAGE & 47.3 & 38.2 & 20.2 & $+18.0$ \\
\cmidrule(lr){1-6}
\multirow{2}{*}{\makecell[l]{AMLNet\\CatBoost}} & Multi-GNN & 74.7 & 21.2 & 22.6 & $-\phantom{0}1.4$ \\
 & GraphSAGE & 74.0 & 20.6 & 23.3 & $-\phantom{0}2.7$ \\
\bottomrule
\end{tabular}

\end{table}
\endgroup

Across all eight settings, the sign follows this replacement balance. Connectivity alone is insufficient for direct review: an added account improves the final queue only when it replaces a less valuable one, and whether it does is a property of the list being replaced rather than an intrinsic property of the account.


\subsection{Capacity-Matched Score Combination (RQ4)}
\label{sec:rq4}

Figure~\ref{fig:score-combination} compares account policies. At $K=100$, the ten-run mean upstream weight $\alpha$ for Multi-GNN/GraphSAGE is 1.00/0.63 for AMLWorld CLeAR and 0.12/0.15 for AMLWorld CatBoost; on AMLNet it is 0.63/0.62 for CLeAR and 0.95/0.98 for CatBoost. CLeAR weights span 0.1--1.0 on AMLNet ($\alpha=1$: three/four runs), 0.2--1.0 for AMLWorld GraphSAGE, and 1.0 for all ten AMLWorld Multi-GNN runs.

For AMLWorld CLeAR with Multi-GNN, validation preserves the upstream count at $K=100$; mean $\alpha$ falls to 0.08 at the pooled crossover $K=3{,}916$, where the fitted order adds 24.7 positives. Lower-yield orders can admit GNN influence from the first capacity. Equal averaging remains slightly better near the AMLNet crossover, so no fixed mixture dominates. Capacity-matched validation, not crossing alone, controls queue influence from held-out yield while the GNN ranks money flows (Table~\ref{tab:capacity-guide}).

\section{Transfer to the HOFINET Benchmark}
\label{sec:hofinet-transfer}

HOFINET~\cite{aihub2026financial} is a synthetic benchmark generated to reflect patterns in real financial activity. It contains 4,732,130 transactions among 452,816 accounts, of which 14,490 transactions carry the source anomaly label. We project that label to both transaction endpoints; a positive account is therefore an endpoint of a dataset-labeled anomalous transaction, not confirmed money-laundering or suspicious-transaction-report ground truth. The account-disjoint test partition contains 89,349 accounts and 511 positive accounts.

The frozen gate and CLeAR selector produce a 5,123-account test pool containing all 511 positives. We retain the bounded one-hop evaluation with ten distinct CLeAR--GNN pairs per architecture.

\begin{table}[h]
  \captionsetup{justification=raggedright,singlelinecheck=false,skip=4pt}
  \caption{HOFINET transfer at $K=100$ from ten selector--GNN pairs; Combined is validation-fitted.}
  \label{tab:hofinet-transfer}
  \centering
  \small
  \setlength{\tabcolsep}{2.0pt}
  \ifdefined\KoreanPaper
  \def\HofiPositive{양성 계좌}
  \def\HofiReplacement{계좌 교체}
  \def\HofiUp{선별}
  \def\HofiRaw{GNN}
  \def\HofiFitted{검증 결합}
  \def\HofiKept{유지}
  \def\HofiAdded{추가}
  \def\HofiRemoved{제거}
  \def\HofiNoCross{$K=5{,}123$까지 지속적 역전 없음.}
\else
  \def\HofiPositive{Positive Accounts}
  \def\HofiReplacement{Replacement}
  \def\HofiUp{Upstream}
  \def\HofiRaw{Raw GNN}
  \def\HofiFitted{Fitted}
  \def\HofiKept{Retained}
  \def\HofiAdded{Added}
  \def\HofiRemoved{Removed}
  \def\HofiNoCross{No sustained crossover through $K=5{,}123$.}
\fi
\begin{tabular}{@{}ccccccc@{}}
\toprule
& \multicolumn{3}{c}{\HofiPositive} & \multicolumn{3}{c}{\HofiReplacement} \\
\cmidrule(lr){2-4}\cmidrule(l){5-7}
GNN & \HofiUp & \HofiRaw & \HofiFitted & \HofiKept & \HofiAdded & \HofiRemoved \\
\midrule
GraphSAGE & 90.5 & 84.2 & 90.5 & 45.6 & 38.6 & 44.9 \\
Multi-GNN & 90.5 & 83.1 & 90.5 & 49.8 & 33.3 & 40.7 \\
\midrule
\multicolumn{7}{@{}l@{}}{\HofiNoCross} \\
\bottomrule
\end{tabular}

\end{table}

At $K=100$, raw GraphSAGE and Multi-GNN ordering loses 6.3 and 7.4 positive accounts per 100 reviews (Table~\ref{tab:hofinet-transfer}); both directions remain significant after Holm-corrected exact sign tests. Positive neighbors are found, but fewer are added than upstream positives are removed, reproducing the main replacement mechanism.

The independently trained transaction decoders move oppositely: GraphSAGE and Multi-GNN improve transaction average precision by 0.203 and 0.275 in all ten pairs. HOFINET therefore reproduces the central separation on a different synthetic surface, not only the account loss. At $K=100$, validation selects the upstream order in all ten runs per GNN ($\alpha=1$), retaining its 90.5-positive mean while the GNN ranks transactions. This is capacity-specific: at $K=25$, mean $\alpha$ is 0.90 for Multi-GNN and 0.75 for GraphSAGE.

HOFINET also supplies a ceiling case: the pool contains all 511 test positives, graph construction adds none from outside it, and no sustained crossover appears through $K=5{,}123$. Complete coverage caps the full-pool total but does not rule out an intermediate crossover, so the absence remains specific to this frozen surface.

No HOFINET path is positive under the prespecified two-transfer definition, so path average precision is undefined. Transaction labels still support ranking, replacement, coverage, and validation-controlled account-queue preservation.

\Needspace{8\baselineskip}
\section{Conclusion}
\label{sec:conclusion}

\methodname{} shows that tracing money and selecting accounts are complementary decisions with separable outcomes. Raw GNN ordering can improve a lower-yield order yet damage a higher-yield one when added positives do not offset those displaced; this cost declines as the stronger prefix is exhausted. HOFINET reproduces this separation on another synthetic surface. Account selection sets the starting surface; capacity-matched validation admits queue changes only where held-out yield supports them, while the GNN ranks money-flow evidence. Institutions can apply this comparison to their own held-out confirmed cases at planned capacity; our reported yields use synthetic transaction-endpoint labels.

\balance
\bibliographystyle{ACM-Reference-Format}
\bibliography{references}

\end{document}